A Potential Aid in the Target Selection for the Comet Interceptor Mission
E. Vigren[1], A. I. Eriksson[1], N. J. T. Edberg[1], C. Snodgrass[2]
[1] Swedish Institute of Space Physics, Uppsala, Sweden, E-mail: erik.vigren@irfu.se
[2] Institute for Astronomy, University of Edinburgh, Blackford Hill, Edinburgh, UK

ABSTRACT

The upcoming Comet Interceptor mission involves a parking phase around the Sun-Earth $L_2$ point before transferring to intercept the orbit of a long period comet, interstellar object or a back-up target in the form of a short-period comet. The target is not certain to be known before the launch in 2029. During the parking phase there may thus arise a scenario wherein a decision needs to be taken of whether to go for a particular comet or whether to discard that option in the hope that a better target will appear within a reasonable time frame later on. We present an expectation value-based formalism that could aid in the associated decision making provided that outlined requirements for its implementation exist.

1. INTRODUCTION

The European Space Agency (ESA) Comet Interceptor (CI) mission (Snodgrass & Jones, 2019, see also https://www.cosmos.esa.int/web/comet-interceptor/home) is planned for launch in 2029. Its goal is to make a flyby of (and thereby study) a dynamically new long period comet (LPC) or an interstellar object (see e.g., Meech et al., 2017; Hoover et al., 2022) while it is passing through the inner solar system. The fact that the target comet possibly will be unknown by the time of launch, makes the mission unique. The mission is designed as such that the spacecraft will be held waiting around Lagrange point 2 of the Sun-Earth system until sent off to intercept the orbit of a target comet by then discovered as recently as within the last few years, either pre- or post launch. The maximum mission duration, meaning the maximum time between launch and target interception, is currently set to 6 years (Jones et al., 2023). The search for potential targets will be carried out by powerful ground based facilities like the Vera C. Rubin Observatory Legacy Survey of Space and Time (LSST) currently under construction in Chile (Ivezić et al., 2019, see also https://www.lsst.org).

Certain requirements need to be fulfilled for a target comet to be considered suitable. For instance, the interception of the comet should not exceed the delta-v budget, must occur within a heliocentric distance range of 0.9-1.2 astronomical units and not occur on the opposite side of the sun as viewed from Earth. There is also, in principle, a limit for the duration of the parking phase and it may happen that a decision needs to be taken to rather go for a back-up target in the form of a short period comet (see e.g., Schwamb et al., 2020).

In this work we touch upon a question that may or may not be faced during the course of the mission; *is it best to go for this particular target comet or to wait and hope that a better one appears later on*? We start in Section 2 by going through requirements for our formalism to possibly be implemented. The four basic ones are:

i) A grading system making it possible to grade recently discovered potential target comets with a value on a continuous scale.

ii) An idea of with what frequency, $\nu$, potential target comets can be expected to appear.

iii) A probability density function telling by which probability a new potential target comet will have a value appearing in any given sub-interval of the scale.

iv) A set deadline for when to stop consider later on observed comets as potential targets.

In Section 3 we detail *the delicate question that may appear* during the mission and go through a formalism to tackle it provided that the aforementioned requirements are met. In Section 4 we show examples of guiding criteria for a few simple probability density functions. We present there also an example of how the formalism can be utilized while invoking relevant distribution functions and a specified grading criteria. These are examples only to demonstrate the formalism, as defining a grading system for the actual Comet Interceptor mission is outside the scope of this study. Concluding remarks are provided in Section 5.

## 2. REQUIREMENTS FOR IMPLEMENTATION

For the tool presented in this work to be of potential use it is required that by the time of implementation there must exist criteria for grading a newly observed target comet on a continuous (0,1)-scale. In principle the formalism can be modified as to work for a scale with different limits, but here we stick to a scale between 0 and 1 as it is without loss of generality. Any LPC with an orbit which can safely be intercepted within the delta-v budget and fulfill other relevant requirements can be taken as appearing on the scale and hence assigned a value, $c$, from the interval (0,1). The best short-period back-up target should ideally be given a value on the same scale (even though it may be needed to apply completely different grading criteria). The orbital parameters of a recently observed LPC can be combined with sophisticated dynamical calculations as to find an optimal interception point with an associated flyby relative velocity (difference between the spacecraft velocity vector and the comet velocity vector) and flyby solar aspect angle (angle between the relative velocity vector and the comet-sun line). In Section 4 we consider as an example a grading system wherein targets associated with a lower magnitude of the flyby relative velocity and with a closer to 90° flyby solar aspect angle are prioritized. We wish to stress, however, that the grading system used in Section 4 is merely one example out of many of how comets can be graded in a quantitative way. To highlight the complexity of the choice it should first of all be mentioned that depending on mission design a very low activity can render poor signal level in various in situ measurements while a high activity and particularly a high dust loss rate can endanger onboard instrumentation (Fulle et al., 2023; Marschall et al., 2022). A high activity target can be viewed as preferable; while the hazard due to increased dust flux can be mitigated by an increased flyby distance the possibility to compensate a low activity by a closer flyby is limited due to finite maximum rotation speed of the mirrors that follow the comet and also due to targeting errors. Of further concern, the earth-sun-comet angle affects the communication and the possibility to make simultaneous remote sensing observations from ground or satellite-based telescopes (see e.g., Meech et al., 2005 and Snodgrass et al., 2017 for supporting ground based campaigns associated with the earlier cometary missions Deep Impact and Rosetta, respectively). Naturally, in addition to risk assessment and encounter geometry there is also a scientific priority involved in picking the ideal target; for instance, a dynamically new comet is prioritized over a returning one (Snodgrass & Jones, 2019).

When the grading system is set a next question to ask is the following: at what frequency, $\nu$, can be expected the discovery of new target comets fulfilling the requirements to be considered feasible? The actual relevant number is the number of comets reachable by Comet Interceptor during its time of operations (i.e. with perihelion/ecliptic crossing during the lifetime of the mission and detected early enough for comet interceptor to catch it). Sánchez et al. (2021) estimate that some 2-3 LPCs per year should be expected to pass through the reachable heliocentric distance range and note that the fraction accessible for CI will depend on features in the spacecraft- and mission design. Current mission design assumptions give a conservative estimate of an 80% probability of there being at least one accessible target over the 6-year lifetime of the mission (Jones et al. 2023). Estimates of $\nu$ will presumably be refined over the years leading up to the launch thanks in part to improved statistics of LPCs. The latter may aid also in the construction of a relevant probability density function, $f(x)$, allowing to answer the question with what probability a newly discovered feasible target comet would appear within a certain sub-interval of the constructed grading scale.

The final parameter needed for implementation of the formalism presented in this work is a set deadline. We shall let $t$ denote the time remaining until the deadline, which we in turn treat as meaning that any long period comet discovered afterwards cannot be considered a suitable target. At this point (if reached) must be accepted to go for any of the already observed long period comets that are still deemed reachable (may be none) or to go for a short period back-up target. In practice, the relevant threshold is the flyby date, not the detection date. With a maximum mission duration of 6 years, a comet detected 4 years after launch cannot be reached if the transfer time is 3 years unless the mission duration is extended (the transfer time is expected to fall within the range of 6 months up to 4 years). Strictly respecting a maximum 6-year mission duration requires modification of the formalism presented in Section 3; in particular, the parameter $v$ should then not be treated as constant in time considering the fact that the time (after launch) of discovery of an LPC will affect whether or not its associated transfer time is acceptable. By considering the deadline as a threshold for the detection date (rather than the flyby date) we avoid the complexity of having to invoke a time dependent $v$.

## 3. DELICATE QUESTION AND GENERAL SOLUTION

Assume that a long period comet, $X$, is discovered, studied and tracked and assigned the value $c$ on the (0,1)-scale after it has been worked out how to best intercept its orbit. The calculated interception trajectory is set to start at the point when time $t$ *remains* to the deadline (according to the current baseline the go for decision needs to be given 6 months before departure from $L_2$). Assume that no better long period comet is observed until a decision must be made (6 months prior to the time of transfer to encounter). Given $c, t, v$ and a probability density function $f(x)$ (for which $f(x) = 0$ whenever $x \notin [0,1]$) we shall provide formalism to tell whether it is correct from an expectation value point of view to go for comet $X$ or not. Before doing so a few points are worthwhile clarifying:

- We assume that $X$ at time $t$ is the still reachable comet with the highest grading.
- We assume that $t$ is positive and that it represents the time to the deadline which we here treat as a discovery threshold date meaning that an LPC discovered afterwards cannot be considered a feasible target. As outlined in Section 2, respecting the nominal maximum mission duration of 6 years requires the utilization of a time dependent $v$. Such can be implemented in the formalism below but in-depth reasoning of the functional form (the time dependence) is beyond the scope of the present work.
- We assume that if $X$ is not chosen as target by time $t$, it is no longer considered as an option.

The general guiding rule is described as follows: *go for X in case its value, c, exceeds the expectation value of the hypothetical target comet one would end up with if one were to start with no comet at t and proceed with analogous strategy until the deadline for considering new observations is reached.*

We can turn the problem into discrete form by dividing the available time $t$ into $N$ bins of equal duration and make use of backward induction (see e.g., Hill 2009). We shall consider a division into bins of so short duration that the probability for detection of multiple candidate targets within the same time bin becomes negligible. We let the time bin in connection with the deadline run between the discrete time points $t_1$ and $t_0$, where $t_0 = 0$ marks the point when the deadline is reached. The expectation value for a hypothetical comet appearing in this interval is given by

$$e_1 = \frac{vt}{N} \int_0^1 x f(x) dx \qquad (1)$$

where $t/N$ is the duration of the time bin, $vt/N$ is the probability that a comet appears during this time interval, and the integral represents the expectation value for such a comet since $f(x)$ is a probability density function over the interval $0 \leq x \leq 1$. By the strategy a comet is to be discarded at $t_1$ if its given value does not exceed $e_1$. When at $t_2$ (further from the deadline than $t_1$) a comet is to be selected only if its value exceeds

$$e_2 = \left(\frac{vt}{N}\right)\left(1 - \int_0^{e_1} f(x)dx\right)\left(\frac{\int_{e_1}^1 xf(x)dx}{\int_{e_1}^1 f(x)dx}\right) + \left(\left[1 - \frac{vt}{N}\right] + \frac{vt}{N}\int_0^{e_1} f(x)dx\right)e_1 \qquad (2)$$

This requires explanation. The first term contains three factors, namely i) the probability that a comet appears during the time interval $t_2$ to $t_1$, ii) the probability that the comet has a value exceeding $e_1$, which is required for selection, and iii) the expectation value of such a comet. The second term covers the two possible strategically correct ways for proceeding to $t_1$ from which point the expectation value is $e_1$ (last factor). The first possible way involves the scenario that no comet appears during the time interval $t_2$ to $t_1$ (probability given by the bracketed expression) and the second way is that a comet appears whose value does not exceed $e_1$. By similar reasoning follows that in general for $k \geq 2$

$$e_k = \left(\frac{vt}{N}\right)\left(1 - \int_0^{e_{k-1}} f(x)dx\right)\left(\frac{\int_{e_{k-1}}^1 xf(x)dx}{\int_{e_{k-1}}^1 f(x)dx}\right) + \left(\left[1 - \frac{vt}{N}\right] + \frac{vt}{N}\int_0^{e_{k-1}} f(x)dx\right)e_{k-1} \quad (3)$$

In the limit of large $N$ the recursively calculated $e_N$ approaches the value to which $c$ should be compared. Only if $c \geq e_N$ is the selection of the target justified from an expectation value point of view.

## 4. EXAMPLE RESULTS AND DISCUSSION

*4.1. Guidelines for some simple probability density functions*

Figure 1a shows four different probability density functions and Fig. 1b shows the corresponding guiding graphs of $c$ value to settle for as a function of the dimensionless quantity $vt$. The probability density functions in question are $f_A(x) = 1$ (black), $f_B(x) = 2x$ (blue), $f_C(x) = 2 - 2x$ (red) and $f_D(x) = 6(x - x^2)$ (magenta). The data points (diamonds) in Fig. 1b were generated through the numerical recursive approach with use of $N = 10^5$ and the $c$ values to settle for are seen to be steadily decreasing with decreasing value of $vt$, which is as expected. Curiously, in the case of a *uniform* probability distribution, as in $f_A(x)$, the criterion for the $c$ value to settle for reduces to $c \geq vt/(vt + 2)$. This can be realized by first noticing that Eq. (3) in the case of a uniform probability distribution after some algebra reduces to the expression $(e_k - e_{k-1})/(vt/N) = (1/2)(1 - e_{k-1})^2$. Then, if introducing $z = vt$, $dz = vt/N$, $e_{k-1} = g(z)$ and $e_k = g(z + dz)$ is seen that as $N \to \infty$ the relation can be liken with the separable differential equation $g'(z) = (1/2)(1 - g(z))^2$, which when combined with the boundary condition $g(0) = 0$, has the solution $g(z) = z/(z + 2)$. That a seemingly tedious problem is associated with such a simple solution is not uncommon in the theory of optimal stopping (c.f., Hill 2009). For the other considered probability density functions in Fig. 1a, we have not yet been able to establish simple guiding relations. This is not a prioritized endeavor, as it is hard to imagine how a constructed grading criteria in combination with sophisticated treatment of LPC statistics is to render probability density functions as trivial as any of the four considered in Fig. 1a. Similar reasoning justifies why we limit $vt$ to values $\leq 14$ in Fig. 1b.

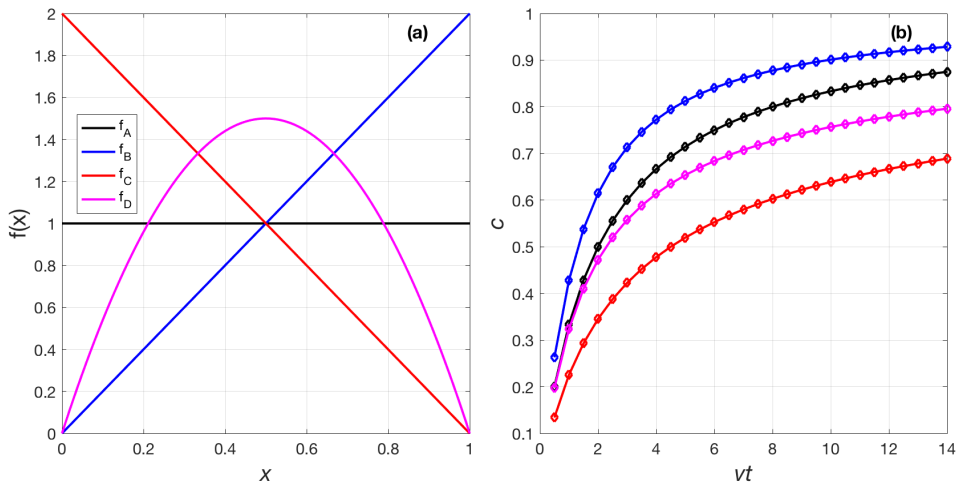

**Figure 1**: (a) Four examples of probability density functions $f_A(x) = 1$ (in black), $f_B(x) = 2x$ (blue), $f_C(x) = 2 - 2x$ (red) and $f_D(x) = 6(x - x^2)$ (magenta) and (b) associated graphs showing what minimum value, $c$, to settle for as a function of the product $vt$.

*4.2. An example invoking available LPC statistics and specified grading criteria*

It is instructive to also study a more involved example. We shall in the following pay respect to relevant LPC statistics available in Jones et al. (2023) as well as in the Definition Study Report of the

Comet Interceptor Mission (see https://www.cosmos.esa.int/web/comet-interceptor/home) and consider a grading system wherein targets are graded based both on the projected encounter relative velocity and the flyby solar aspect angle. We use in particular presented histograms over the modulus of the encounter relative velocity, $u$, and the flyby solar aspect angle, $\xi$, for simulated encounters obtained from the population of LPCs. We have digitized these and reproduce by dots (at bin centers) in Figs. 2a and 2b scaled down values (bin heights) over certain restricted intervals. The restrictions are made as to limit potentially considered target comets to ones with $u \leq 70$ km s$^{-1}$ and with $45° \leq \xi \leq 135°$ (the restrictions, justified below, are estimated to combined remove ~15% of the population subject to no restrictions on $u$ or $\xi$). Shown in Figs. 2a and 2b are also fits to high order polynomial functions adjusted as to represent probability density functions over the given intervals. The associated cumulative distribution functions are shown in Figs. 2c and 2d, respectively.

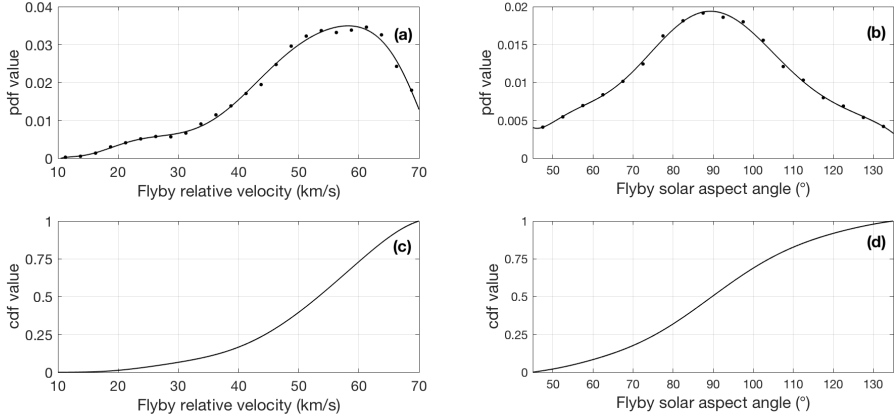

**Figure 2**: The dots in panels (a) and (b) represent scaled down values of bin heights presented in histograms over encounter relative velocity and flyby solar aspect angle in Jones et al. (2023) and in the Definition Study Report of the Comet Interceptor Mission. The solid lines are polynomial fits modified to represent probability density functions over the given intervals. Panels (c) and (d) show the associated cumulative distribution functions.

Turning to the subject of grading criteria it is noted that a low value of the modulus of the encounter velocity, $u$, is desirable not only because it would mean a longer duration of measurements. A high flyby velocity means also a greater risk for severe instrument- or spacecraft damage upon strike by cometary dust particles (e.g., Fink et al., 2021; Marschall et al., 2022). Also, a high spacecraft velocity can make impact of cometary coma molecules a strong driver of electron emission from the spacecraft and thereby complicate the analysis/interpretation of *in situ* plasma measurements (e.g., Grard et al., 1988; Johansson et al., 2022). Additionally, a lower velocity is valuable as it makes it easier for rotating mirrors to track the nucleus. While designed for the limiting 70 km/s case, a lower speed means an easier tracking and the possibility of a closer flyby for the same angular velocity of tracking mirrors. For the flyby solar aspect angle, $\xi$, a value near 90° brings optimal illumination conditions for imaging and a requirement of $45° \leq \xi \leq 135°$ is set as driven by sun avoidance for remote sensing, and also spacecraft power needs, as solar panels will be kept edge on to ram direction to minimize dust impacts during flyby. With these requirements in mind, let us, just as an example, assume that we are to grade comets via $((u_{min} + u_{max} - u)/u_{max})(2 \sin^2 \xi - 1)$ where we adapt $u_{min} = 10$ km s$^{-1}$ and $u_{max} = 70$ km s$^{-1}$ and where we restrict 10 km s$^{-1} \leq u \leq 70$ km s$^{-1}$ and $45° \leq \xi \leq 135°$. Assuming independence, sampling $u_i$ and $\xi_i$, while respecting the distribution functions of Figs. 2a and 2b, respectively, can be done by drawing two random numbers, $r_{i_1}$ and $r_{i_2}$, with uniform probability distribution from (0,1), and then interpolate with respect to the cumulative distribution functions of Figs. 2c and 2d, respectively. In Fig. 3a is plotted scaled down bin heights from a histogram based on $2 \times 10^5$ sample points of the form $f_i = ((u_{min} + u_{max} - u_i)/u_{max})(2 \sin^2 \xi_i - 1)$. A fit, $f(x)$, in form of a polynomial with properties of a probability density function is shown by the gray line while Fig. 3b shows the corresponding $c$ value to settle for as a function of $vt$ as determined from the recursive approach outlined in Section 3. We refrain from presenting explicitly the involved polynomial fit functions as to reduce the risk of overselling the results from this exercise, as it is intended mainly as a demonstration of that the formalism can be applied also for non-trivial probability density functions. We remark also that uncertainties in the distribution of cometary properties naturally will propagate into uncertainties in the value of $f(x)$, but deem a proper error analysis to be out of the scope of the present work.

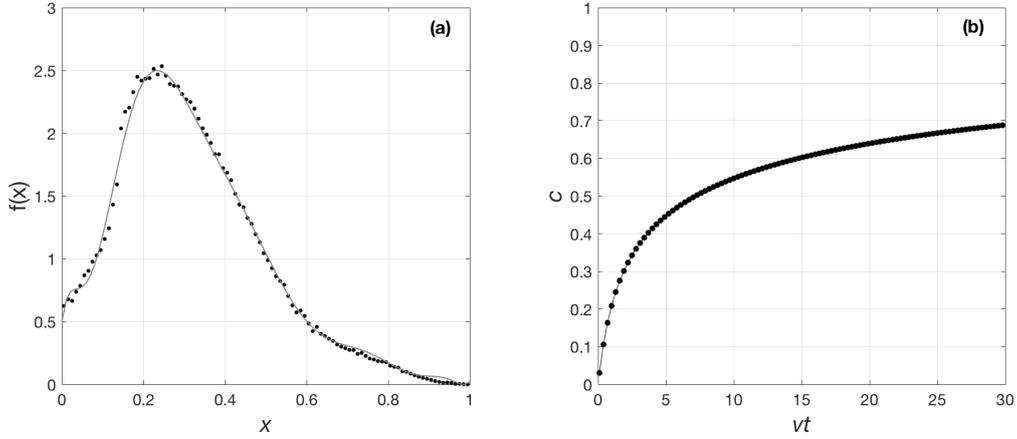

**Figure 3:** The dots in panel (a) relate to a histogram of $((u_{min} + u_{max} - u_i)/u_{max})(2\sin^2\xi_i - 1)$ when sampled over $2\times 10^5$ random points respecting the distributions in Fig. 2a and Fig. 2b. The gray solid line is a polynomial fit adjusted as to also represent a probability density function. In panel (b) is shown the associated $c$ value to settle for as a function $vt$.

5: CONCLUDING REMARKS

We have presented formalism of potential use in the ultimate selection of target for the Comet Interceptor mission should the question emerge of whether it is better to go for an available target or risk waiting (and thereby "loosing it") in hope of that a better target is observed within a reasonable time frame later on. We have outlined what is required for implementation of the formalism and illustrated its potential use through an example. While the presented formalism applies for a scenario where a deadline is set for the latest target discovery date it can in principle be modified to instead encompass a latest flyby date. Such a modification requires "only" the utilization of a time varying frequency parameter $v$ and is important to make when wanting to pay strict respect to the nominal maximum mission duration of 6 years (referring to the time between the launch and the point of interception). It is our hope that the delicate question treated in this work never actually will be faced during the course of the Comet Interceptor mission. Given that the observational surveys are expected to offer *warning times* (time from discovery to the interception opportunity) of typically more than three years (Sánchez et al., 2021), it is in fact improbable that it will. Still, the question *may* appear, and in that event it is good to have strategies based on probability theory at hand. To this end it may be further noted that the probability density functions considered in this work all yield a scenario where at low value of $vt$ the $c$ value to settle for changes quickly. This highlights that the reliability in using our formalism is tightly connected to the reliability/accuracy in the estimate of the frequency of discovery of suitable comets, $v$. Work is already ongoing on constraining $v$ and such efforts will benefit greatly from the LSST once it is running. As a final remark is to be noted that the following of the strategy outlined in this work, even in the scenario of a well-constrained parameter set, is correct only in the average sense. In the isolated case it is not necessarily rewarding to turn down a semi-decent offer.

ACKNOWLEDGEMENT

The work of NJTE was supported by SNSA grant 2021-00047. We are thankful to Andreas Dieckmann for helping out in verifying the correctness of the presented exact condition for $c$ value to settle for in the case of a uniform probability density function. We are also thankful to reviewers of the original manuscript who raised important remarks, thereby contributing to improving our work.